\documentclass[12pt]{article}
\usepackage{colordvi}
\usepackage{latexsym}
\usepackage{epsfig}
\begin{document}

\begin{center}
{\Large \textbf{Long-lived Charginos in the  Focus-point Region\\[0.3cm]
of the MSSM Parameter Space}}

\vspace{10mm}

\large A.~V.~Gladyshev$^{1,2}$, D.~I.~Kazakov$^{1,2}$ and M.~G.~Paucar$^1$

\normalsize
\vspace{5mm}
\textit{$^1$ Bogoliubov Laboratory of Theoretical Physics,\\
Joint Institute for Nuclear Research,\\
141980, 6 Joliot-Curie, Dubna, Moscow Region, Russian Federation}

\vspace{1mm}
 \textit{$^2$ Institute for Theoretical and Experimental Physics,\\
 117218, 25 Bolshaya Cheremushkinskaya, Moscow, Russian Federation}
\end{center}
\vspace{5mm}
\begin{abstract}
We analyse the possibility to get light long-lived charginos
within the framework of the MSSM with gravity mediated SUSY
breaking. We find out that this possibility can be realized in
the so-called focus-point region of parameter space. The mass
degeneracy of higgsino-like chargino and two higgsino-like
neutralinos is the necessary condition for a long lifetime. It
requires the fine-tuning of parameters, but being a single
additional constraint in the whole parameter space it can be
fulfilled in the Constrained MSSM along the border line where
radiative electroweak symmetry breaking fails. In a narrow band
close to the border line the charginos are long-lived particles. 
The cross-sections of their production and co-production at the 
LHC via electroweak interaction reach a few tenth of pb.
\end{abstract}

\section{Introduction}

An expected discovery of SUSY at the LHC is based on the MSSM
predictions with a wide range of possibilities and uncertainties~
\cite{MSSM1,MSSM2,MSSM3}. The usual pattern is that one has
heavy short-lived particles which are created in strong and weak
interactions and then decay (almost immediately) into
Standard Model particles. These processes have a natural Standard Model
background and their observation is rather sophisticated.

There are, however, some exotic cases, which are more natural
within the gauge mediation scenario but may also be realized in
mSUGRA when SUSY particles are long-lived. In the latter case,
this usually takes place when their masses are degenerate with
the LSP mass and the decay rate therefore is suppressed. This
degeneracy in its turn requires precise fine-tuning of
parameters of the mSUGRA parameter space. However, having at least
four parameters ($m_0,m_{1/2},A_0$ and $\tan\beta$) one has
enough freedom to satisfy this single additional requirement.

In what follows we analyse the parameter space of the MSSM with
gravity mediated SUSY breaking
with the aim to find the regions where the above-mentioned
degeneracy can take place and hence the long-lived superpartners
might exist. In our previous papers we considered two of such
possibilities: long-lived staus which live in the so-called
coannihilation region~\cite{us1,us2}, and long-lived stops which
live at the edge of the bulk region near the Higgs limit border
line~\cite{us3,us4}.

Here we explore another region of parameter space which is a
narrow band along the line where the radiative electroweak
symmetry breaking fails.  On the border of this region the Higgs
mixing parameter $\mu$, which is determined from the requirement
of electroweak symmetry breaking via radiative corrections, tends
to zero. This leads to existence of light and degenerate states:
the second chargino and two neutralinos, all of them being
essentially higgsinos. Below we discuss their properties and
possible creation at the LHC.

\section{Constraints on the  MSSM Parameter Space for Large
        Values of $A$}

In what follows we consider the MSSM with gravity mediated
supersymmetry breaking and the universal soft terms. We thus have
the parameter space defined by $m_0, m_{1/2},$ $A, \tan\beta$ and
we take the sign of $\mu$ to be positive motivated by contribution
to the anomalous magnetic moment of
muon~\cite{amu1,amu2,amu3}. Imposing the
constraints like: i) the gauge couplings
unification~\cite{unif1,unif2,unif3}, ii) neutrality of the
LSP~\cite{lsp1,lsp2}, iii) the Higgs boson and SUSY mass
experimental limits~\cite{higgslimits1,higgslimits2,higgslimits3},
iv) radiative electroweak symmetry breaking~\cite{ewsb1,ewsb2}, we
get the allowed region of parameter space of the CMSSM. Projected
onto the $m_0,m_{1/2}$ plane this region depends on the values of
$\tan\beta$ and $A$.

We are interested in chargino and neutralino masses. The mass
terms are nondiagonal and look like
\begin{equation}
{\cal L}_{Gaugino-Higgsino}=
 -\frac{1}{2}M_3\bar{\lambda}_a\lambda_a
 -\frac{1}{2}\bar{\chi}M^{(0)}\chi -(\bar{\psi}M^{(c)}\psi + h.c.),
\end{equation}
where $\lambda_a , a=1,2,\ldots ,8,$ are the Majorana gluino fields and
\begin{equation}
\chi = \left(\begin{array}{c}\tilde{B}^0 \\ \tilde{W}^3 \\
\tilde{H}^0_1 \\ \tilde{H}^0_2
\end{array}\right), \ \ \ \psi = \left( \begin{array}{c}
\tilde{W}^{+} \\ \tilde{H}^{+}
\end{array}\right)
\end{equation}
are, respectively, the Majorana neutralino and Dirac chargino fields.

At the tree level the neutralino mass matrix is
\begin{equation}
M^{(0)}=\left(
\begin{array}{cccc}
M_1 & 0 & -M_Z\cos\beta \sin_W & M_Z\sin\beta \sin_W \\  0 & M_2 &
M_Z\cos\beta \cos_W   &
-M_Z\sin\beta \cos_W  \\ -M_Z\cos\beta \sin_W & M_Z\cos\beta
\cos_W  & 0 & -\mu \\
M_Z\sin\beta \sin_W & -M_Z\sin\beta \cos_W  & -\mu & 0
\end{array} \right),\label{neut}
\end{equation}
where $\tan\beta = v_2/v_1$ is the ratio of two Higgs  v.e.v.s and
$\sin_W= \sin\theta_W$ is the usual sinus of the weak mixing
angle. The physical neutralino masses $m_{\tilde{\chi}_i^0}$ are
obtained as eigenvalues of this matrix after diagonalization.

For charginos one has
\begin{equation}
M^{(c)}=\left(
\begin{array}{cc}
M_2 & \sqrt{2}M_W\sin\beta \\ \sqrt{2}M_W\cos\beta & \mu
\end{array} \right).\label{char}
\end{equation}

This matrix has two chargino eigenstates
$\tilde{\chi}_{1,2}^{\pm}$  with mass eigenvalues
\begin{eqnarray}
M^2_{1,2}&=&\frac{1}{2}\left[M^2_2+\mu^2+2M^2_W \right. \label{charmass} \\
&\mp& \left. \sqrt{(M^2_2-\mu^2)^2+4M^4_W\cos^22 \beta
+4M^2_W(M^2_2+\mu^2+2M_2\mu \sin 2\beta )}\right].\nonumber
\end{eqnarray}

These matrices obtain radiative corrections which are known in the
leading order~\cite{rad1,rad2}. Typically they are of the order of a few
per cent.

\begin{figure}[bt]
    \epsfxsize=7.5cm \epsffile{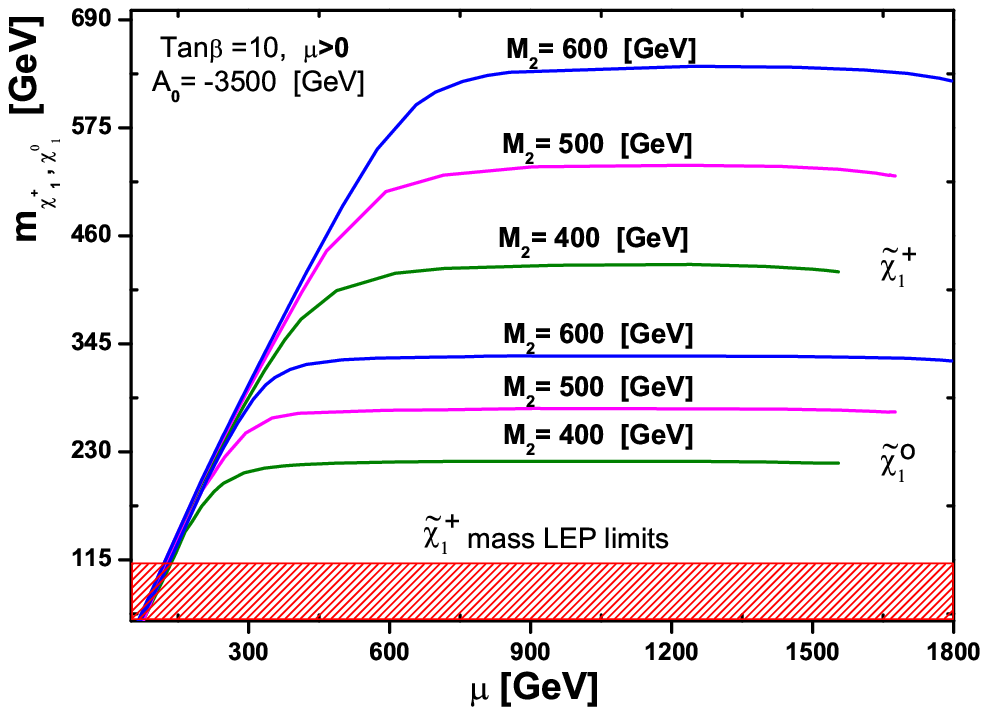}
    \epsfxsize=7.3cm \epsffile{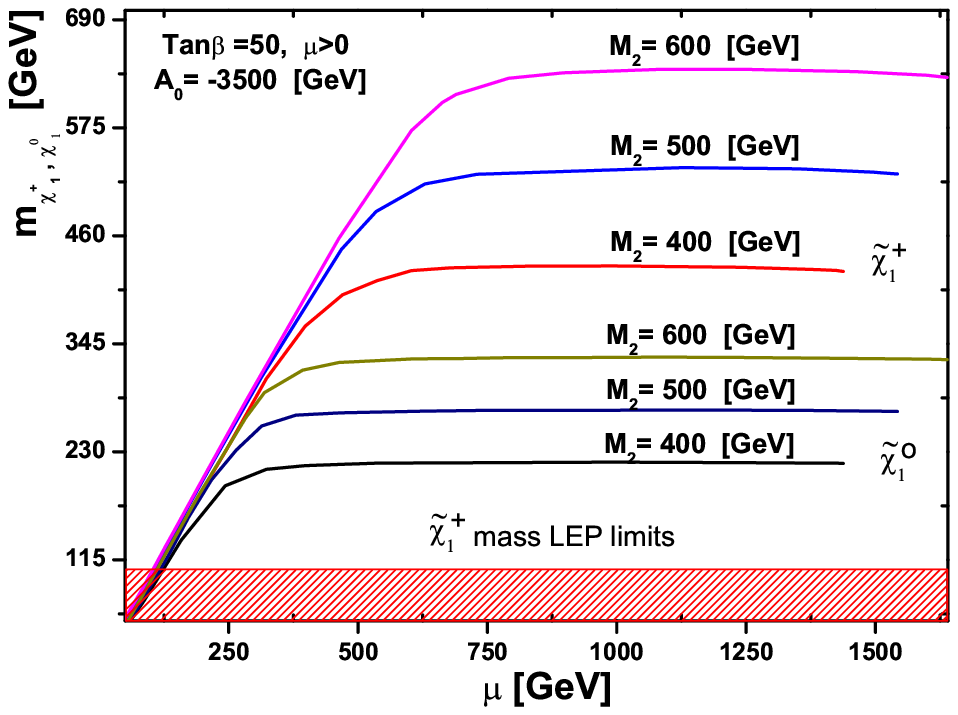}
\caption{The masses of the lightest chargino and neutralino
as functions of $\mu$ for the rest of parameters fixed. The value
of $M_2$ is taken to be 600, 500 and 400 GeV, and
$\tan\beta$=10,50, respectively. Dark (red) lower band shows the
experimental limit on chargino mass
\label{mass}}
\end{figure}

When $\mu$ is small (less than $M_Z$), which takes place
in the focus point region near the border line of radiative
electroweak symmetry breaking, the lightest chargino
($\chi^\pm_1$) and two lightest neutralinos ($\chi^0_{1,2})$ are
almost degenerate and have a mass of the order of $\mu$, which
corresponds to the minus sign in Eq.\ref{charmass}. All of them in
this case are predominantly higgsinos. In Fig.\ref{mass} it is
shown how the mass of the lightest neutralino and the mass of the
lightest chargino depend on $\mu$.

We are interested in the region of small $\mu$ where the three
masses ($m_{\chi^0_1},m_{\chi^0_2},m_{\chi^\pm_1}$) are almost
degenerate. This takes place for any choice of the other
parameters since the tree level formulae weakly depend on them and
the one-loop corrections are usually small. However, since the
value of $\mu$ is not arbitrary in this approach but is taken from
the requirement of electroweak symmetry breaking, one has to find
the region of parameter space where it is small. In Fig.\ref{mass}
this region is just above the chargino LEP limit in the left
bottom corner of the plots. One can see that masses are
degenerate, and there the value of $\mu$ is of the order of
150--200 GeV depending on the value of $\tan\beta$. There is also
a slight dependence on $M_2$ (that is on $m_{1/2}$); however, this
dependence only shows how far we may go along the lines having
masses degenerate. It is clearly seen that the bigger $M_2$ the
larger values of $\mu$ are allowed. The mass of $\chi^0_2$ is not
shown, it almost coincides with the $\chi^\pm_1$ mass.

\begin{figure}[phtb]
    \epsfxsize=7.5cm \epsffile{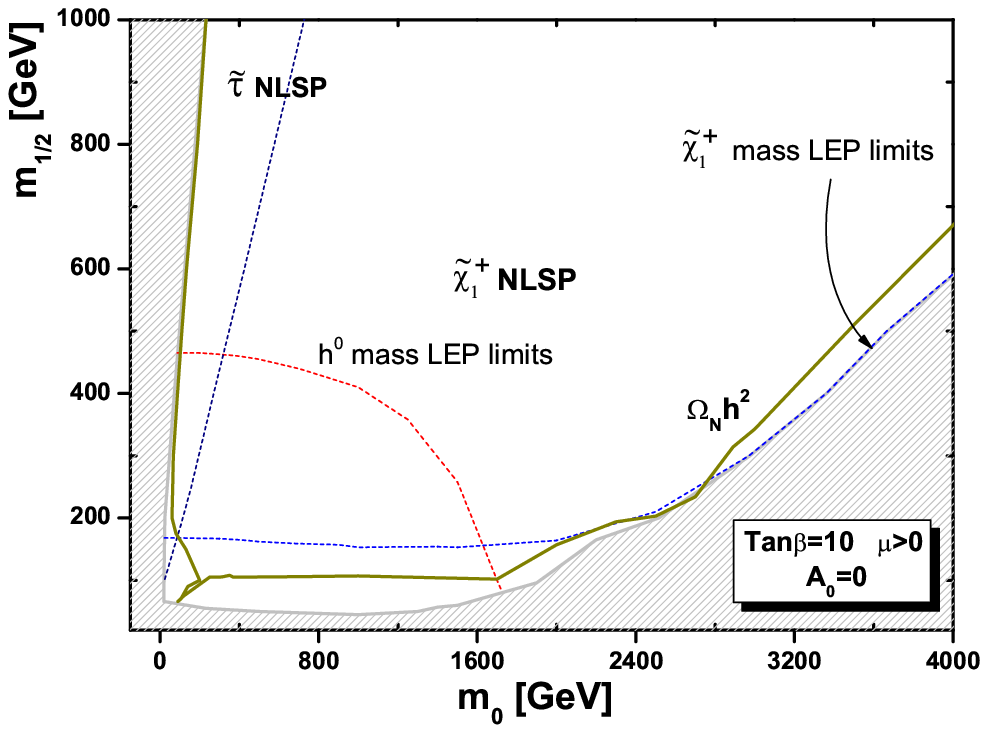}
    \epsfxsize=7.4cm \epsffile{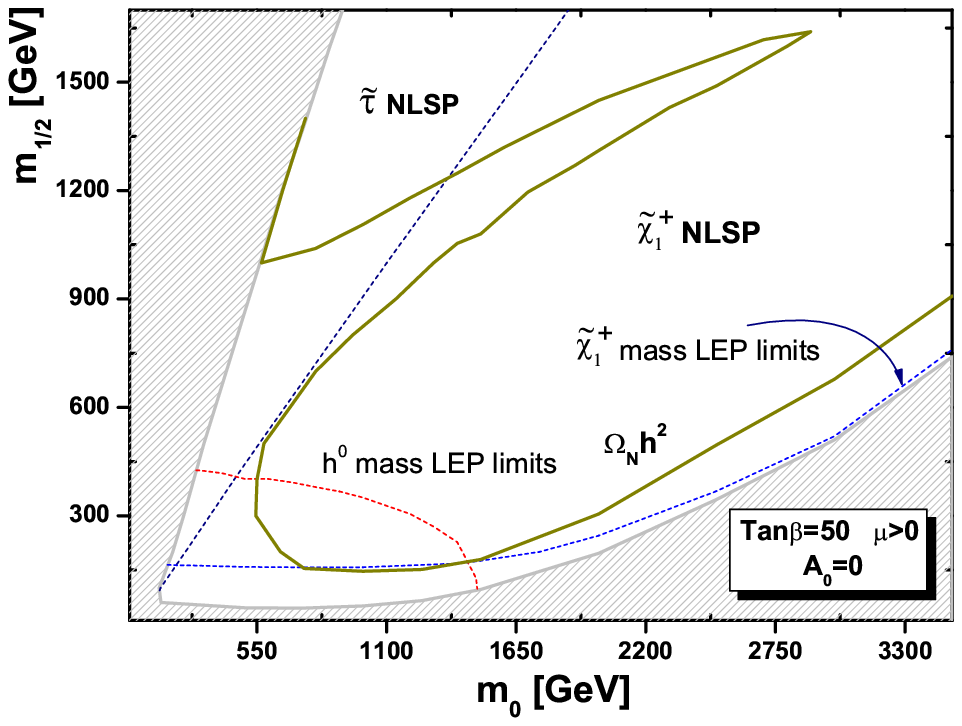}
    \epsfxsize=7.5cm \epsffile{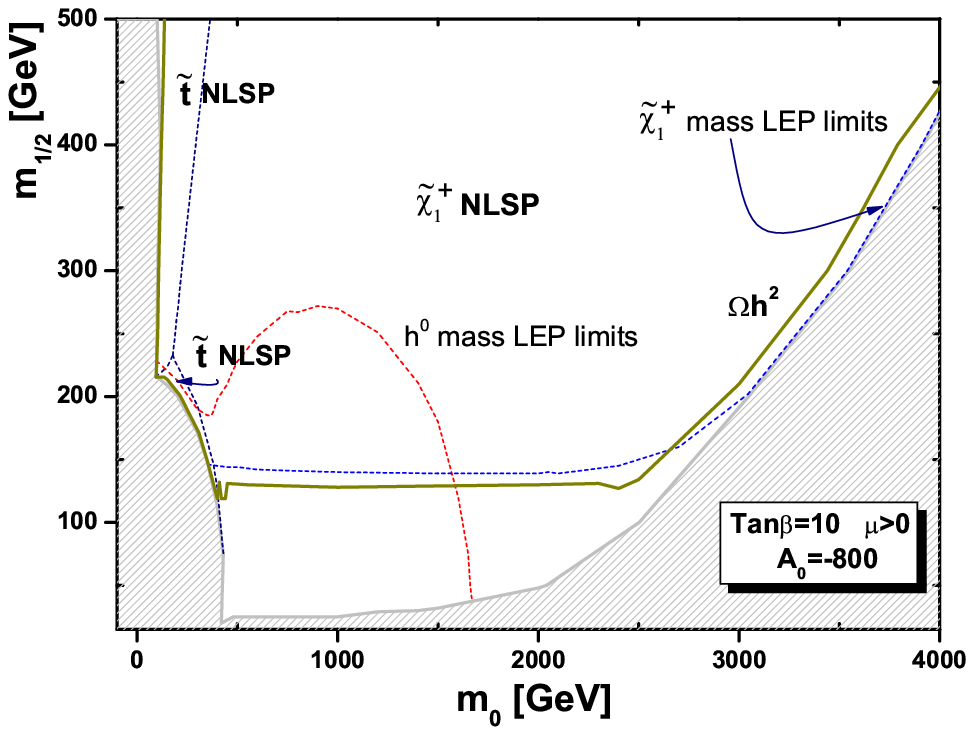}
    \epsfxsize=7.5cm \epsffile{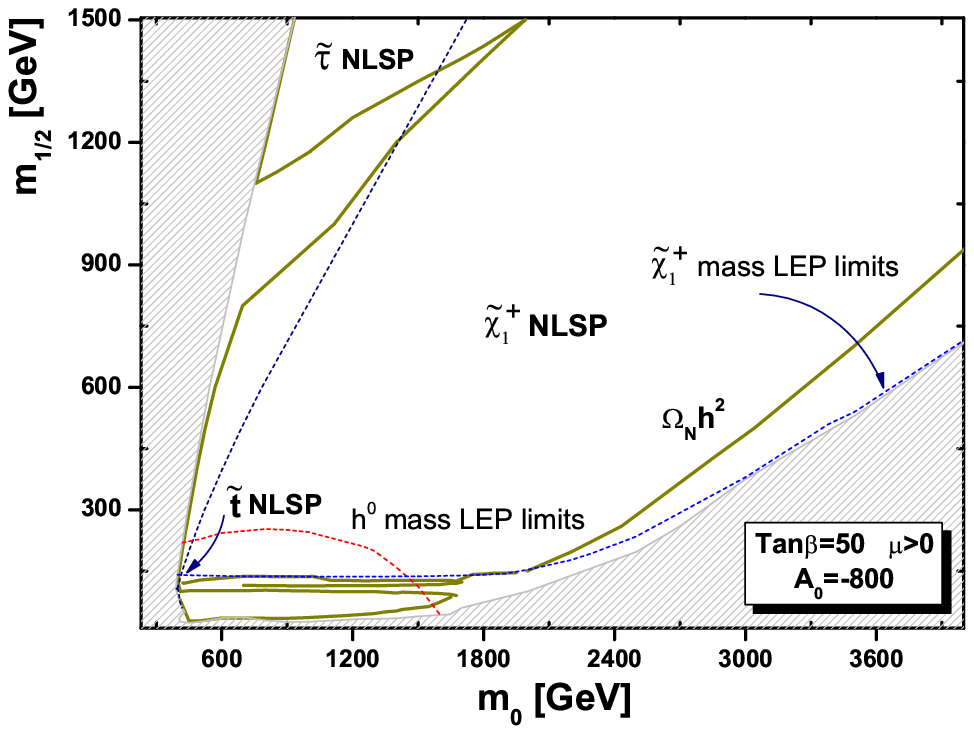}
    \epsfxsize=7.4cm \epsffile{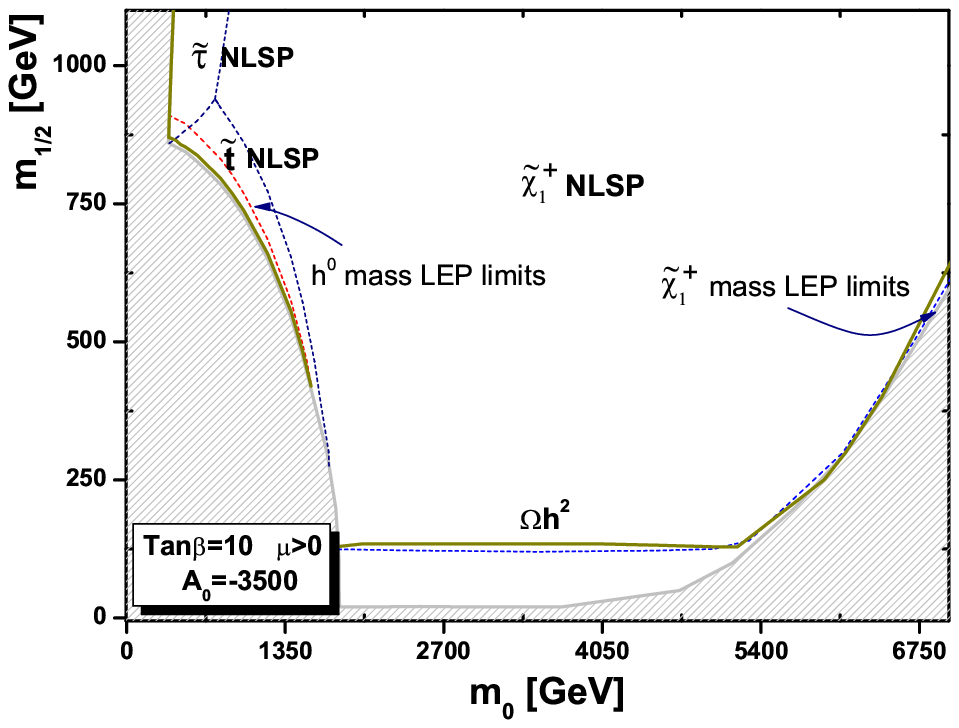} \hspace*{2mm}
    \epsfxsize=7.3cm \epsffile{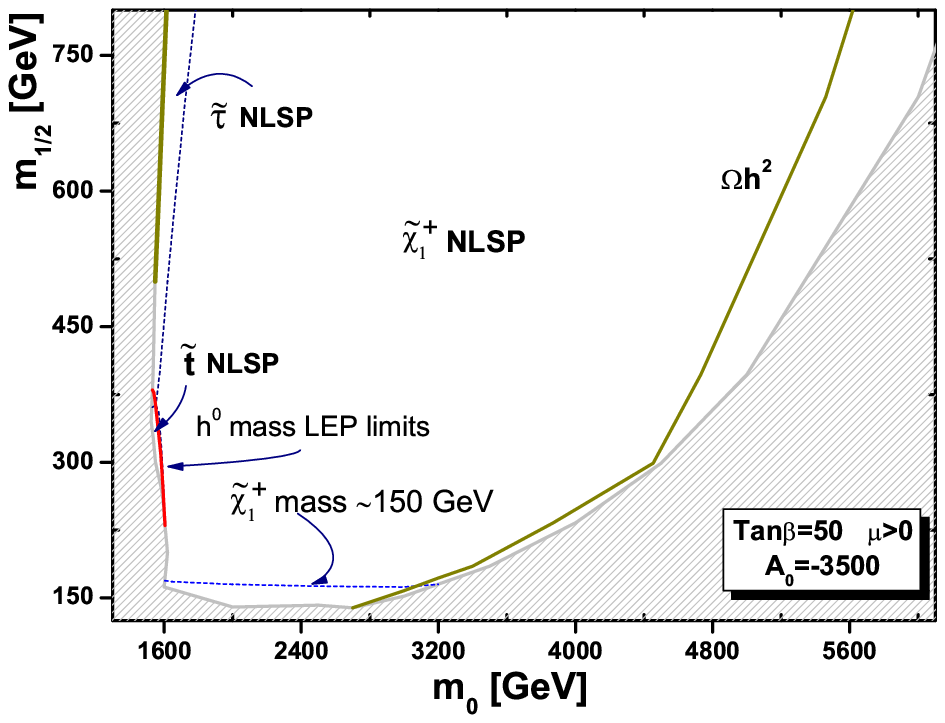}
\caption{Allowed region of the mSUGRA parameter space
for $A_0=0,-800,-3500$ GeV and $\tan\beta=10,50$, respectively.
Dark (blue) areas show theoretically forbidden regions. Along the
narrow green curve the amount of the Dark matter corresponds to
WMAP data~\cite{wmap1,wmap2} $\Omega h^2=0.09\pm 0.04$. Experimental
limits on the Higgs and chargino masses are shown as well.
 \label{regions}}
\end{figure}

One has to check also that the other requirements on the values of the mSUGRA
parameters are satisfied in this region. The well-known
analysis of the CMSSM shows that small
values of $\mu$ are achieved in the so-called focus-point region
along the border line of electroweak symmetry breaking~\cite{fpr1,fpr2,fpr3}. In
Fig.\ref{regions} we show the projection of SUSY parameter space
onto the $m_0,m_{1/2}$ plane  for different values of $A$ and
$\tan\beta$. To calculate it, we use the SuSpect v2.3.4 code
\cite{suspect}\footnote{We have set the following input values at
$M_{W}$; $\alpha_{em}^{-1}=127.934$, $\alpha_{s}=0.1172$,
$m_{t}^{pole}=172.7$ GeV, $m_{b}(m_{b})^{\overline{MS}}=4.25$ GeV,
$m_{\tau}^{pole}=1.777$ GeV.}. The relic density was calculated
with the help of the MicrOmegas package~\cite{micromegas1,micromegas2}. In
the case of light chargino and the second neutralino when they are
almost degenerate with the lightest neutralino, when calculating
the relic density one has to take into account not only the
annihilation of two neutralinos but also the coannihilation of
charginos and neutralinos.

One can see that for small values of $A_0$ the Dark matter line
does not go along the electroweak symmetry breaking border but
deviates from it, thus not allowing the small values of $\mu$. For
large and negative $A_0$, on the contrary, these two lines almost
coincide, the bigger the value of $\tan\beta$ the better.

It should be mentioned that the region near the border line is
very sensitive to the Standard Model parameters; a minor shift in
$\alpha_s$ or $m_t$ and $m_b$ leads to a noticeable change of
the spectrum, as can be seen from comparison of different codes
at~\cite{kraml1,kraml2,kraml3}.

Note that though the region of small $\mu$ looks very fine-tuned
and indeed is very sensitive to all input parameters, still in the
whole four dimensional parameter space (assuming universality) it
swaps a wide area and can be easily reached. The accuracy of
fine-tuning defines the accuracy of degeneracy of the masses and,
hence, the life time of the NLSP which is the lighest chargino.

\section{Phenomenological Consequences of the Light Charginos}

\begin{figure}[ht]
\vspace{0.5cm}
\begin{center}
    \epsfxsize=10cm \epsffile{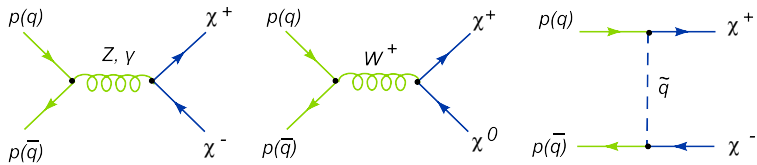}
\caption{Main chargino production diagrams at the LHC}
\label{prod}
\end{center}
\end{figure}

Whence the parameters are chosen in such a  way that one has mass
degeneracy between the lightest chargino and the lightest
neutralino and thus one has a long-lived NLSP. Its mass is typically in the 100
GeV range and  the cross-section of production at the LHC is
considerably high. The main production diagrams are shown in
Fig.\ref{prod}. Since three states are almost degenerate, one has
also co-production which has to be taken into account. This refers
also to the annihilation process that defines the amount of the
Dark matter.

To calculate the production rate, one has to know the spectrum of
the light states and the mixings in the chargino-neutralino sector. In
our case, the lightest chargino and the lightest neutralinos are
almost fully higgsinos. This property defines the preferences in
the interaction pattern. To be specific, we choose several points
in the mSUGRA parameter space which are situated in the focus point
region along the border of electroweak symmetry breaking (see
Fig.\ref{regions}). The values of $\mu$ and $\tan\beta$ are
varied. As we have already mentioned for consistency with the
relic density constraint the advocated scenario prefers large
negative $A_0$ and large $\tan\beta$.

For each point we have calculated the cross-sections of chargino
production and co-production with the lightest and
next-to-lightest neutralino. The main reactions are
$$pp\rightarrow\widetilde{\chi}_{1}^{+}\widetilde{\chi}_{1}^{-}
, \ \ \ pp\rightarrow\widetilde{\chi}_{1}^{+}\widetilde{\chi}_{1}^{0}, \ \ \
pp\rightarrow\widetilde{\chi}_{1}^{+}\widetilde{\chi}_{2}^{0}$$

The tree level processes are shown in Fig.\ref{prod}.
The cross-sections are
calculated with the help of the CALCHEP code~\cite{calchep} with the
MRST parton distribution~\cite{mrst} and are summarized in Tables
\ref{t120}-\ref{t200}. Shown also are the partial cross-sections for
particular quarks inside the proton.

One can see that on average the cross-sections reach a few tenth of
pb and slightly vary with the change of $\tan\beta$. The
cross-sections mainly depend on $\mu$: the bigger the value of $\mu$ the
smaller the cross-section. This is due to the fact that the
cross-section mostly depends on the masses determined mostly by
$\mu$ and mixings.

The values of parameters used in our calculations are the following:
we take $M_1=200$ GeV, and all squark and slepton masses are
around 1000 GeV. The values of $\mu$ are 120, 150 and 200 GeV, and
$\tan\beta$ values are 10, 30, 50.

\begin{table}[phtb]
\begin{center}
\begin{tabular}{|c|c|c|c|}
\hline
Process & $\tan\beta=10$ & $\tan\beta=30$ & $\tan\beta=50$ \\
\hline
& \multicolumn{3}{c|}{$pp\rightarrow\widetilde{\chi}_{1}^{+}\widetilde{\chi}_{1}^{-}$}\\
\cline{2-4}
$\overline{u}u\rightarrow\widetilde{\chi}_{1}^{+}\widetilde{\chi}_{1}^{-}$
& 0.46 & 0.45 & 0.45 \\
$u\overline{u}\rightarrow\widetilde{\chi}_{1}^{+}\widetilde{\chi}_{1}^{-}$
& 0.46 & 0.45 & 0.45 \\
$\overline{d}d\rightarrow\widetilde{\chi}_{1}^{+}\widetilde{\chi}_{1}^{-}$
& 0.20 & 0.19 & 0.19 \\
$d\overline{d}\rightarrow\widetilde{\chi}_{1}^{+}\widetilde{\chi}_{1}^{-}$
& 0.20 & 0.19 & 0.19 \\
\hline
$\sigma_{p_{1}}(pb)$  & 1.32 & 1.28 & 1.28 \\

\hline & \multicolumn{3}{c|}{$pp\rightarrow\widetilde{\chi}_{1}^{+}\widetilde{\chi}_{1}^{0}$}\\
\cline{2-4} $u\overline{d}\rightarrow\widetilde{\chi}_{1}^{+}\widetilde{\chi}_{1}^{0}$
& 0.82 & 0.79 & 0.78 \\
$\overline{d}u\rightarrow\widetilde{\chi}_{1}^{+}\widetilde{\chi}_{1}^{0}$
& 0.82 & 0.79 & 0.78 \\
\hline
$\sigma_{p_{2}}(pb)$  & 1.63 & 1.58 & 1.56 \\

\hline & \multicolumn{3}{c|}{$pp\rightarrow\widetilde{\chi}_{1}^{+}\widetilde{\chi}_{2}^{0}$}\\
\cline{2-4} $u\overline{d}\rightarrow\widetilde{\chi}_{1}^{+}\widetilde{\chi}_{2}^{0}$
& 0.67 & 0.66 & 0.66 \\
$\overline{d}u\rightarrow\widetilde{\chi}_{1}^{+}\widetilde{\chi}_{2}^{0}$
& 0.67 & 0.66 & 0.66 \\
\hline
$\sigma_{p_{3}}(pb)$  & 1.34 & 1.32 & 1.32 \\
\hline
\end{tabular}
\end{center}
\caption{The cross-sections for the chargino production  and
co-production at the LHC in the focus point region for $\mu=120$
GeV and $\tan\beta=10,30,50$\label{t120}}
\end{table}

\begin{table}[phtb]
\begin{center}
\begin{tabular}{|c|c|c|c|}
\hline
Process & $\tan\beta=10$ & $\tan\beta=30$ & $\tan\beta=50$ \\
\hline
& \multicolumn{3}{c|}{$pp\rightarrow\widetilde{\chi}_{1}^{+}\widetilde{\chi}_{1}^{-}$}\\
\cline{2-4}
$\overline{u}u\rightarrow\widetilde{\chi}_{1}^{+}\widetilde{\chi}_{1}^{-}$
& 0.22 & 0.22 & 0.21 \\
$u\overline{u}\rightarrow\widetilde{\chi}_{1}^{+}\widetilde{\chi}_{1}^{-}$
& 0.22 & 0.22 & 0.21 \\
$\overline{d}d\rightarrow\widetilde{\chi}_{1}^{+}\widetilde{\chi}_{1}^{-}$
& 0.09 & 0.09 & 0.09 \\
$d\overline{d}\rightarrow\widetilde{\chi}_{1}^{+}\widetilde{\chi}_{1}^{-}$
& 0.09 & 0.09 & 0.09 \\
\hline
$\sigma_{p_{1}}(pb)$  & 0.62 & 0.61 & 0.60 \\

\hline & \multicolumn{3}{c|}{$pp\rightarrow\widetilde{\chi}_{1}^{+}\widetilde{\chi}_{1}^{0}$}\\
\cline{2-4} $u\overline{d}\rightarrow\widetilde{\chi}_{1}^{+}\widetilde{\chi}_{1}^{0}$
& 0.35 & 0.35 & 0.34 \\
$\overline{d}u\rightarrow\widetilde{\chi}_{1}^{+}\widetilde{\chi}_{1}^{0}$
& 0.35 & 0.35 & 0.34 \\
\hline
$\sigma_{p_{2}}(pb)$  & 0.70 & 0.69 & 0.68 \\

\hline & \multicolumn{3}{c|}{$pp\rightarrow\widetilde{\chi}_{1}^{+}\widetilde{\chi}_{2}^{0}$}\\
\cline{2-4} $u\overline{d}\rightarrow\widetilde{\chi}_{1}^{+}\widetilde{\chi}_{2}^{0}$
& 0.32 & 0.31 & 0.31 \\
$\overline{d}u\rightarrow\widetilde{\chi}_{1}^{+}\widetilde{\chi}_{2}^{0}$
& 0.32 & 0.31 & 0.31 \\
\hline
$\sigma_{p_{3}}(pb)$  & 0.64 & 0.63 & 0.62 \\
\hline
\end{tabular}
\end{center}
\caption{The cross-sections for the chargino production  and
co-production at the LHC in the focus point region for $\mu=150$
GeV and $\tan\beta=10,30,50$\label{t150}}
\end{table}

\begin{table}[ptb]
\begin{center}
\begin{tabular}{|c|c|c|c|}
\hline
Process & $\tan\beta=10$ & $\tan\beta=30$ & $\tan\beta=50$ \\
\hline
& \multicolumn{3}{c|}{$pp\rightarrow\widetilde{\chi}_{1}^{+}\widetilde{\chi}_{1}^{-}$}\\
\cline{2-4}
$\overline{u}u\rightarrow\widetilde{\chi}_{1}^{+}\widetilde{\chi}_{1}^{-}$
& 0.09 & 0.08 & 0.08 \\
$u\overline{u}\rightarrow\widetilde{\chi}_{1}^{+}\widetilde{\chi}_{1}^{-}$
& 0.09 & 0.08 & 0.08 \\
$\overline{d}d\rightarrow\widetilde{\chi}_{1}^{+}\widetilde{\chi}_{1}^{-}$
& 0.03 & 0.03 & 0.03 \\
$d\overline{d}\rightarrow\widetilde{\chi}_{1}^{+}\widetilde{\chi}_{1}^{-}$
& 0.03 & 0.03 & 0.03 \\
\hline
$\sigma_{p_{1}}(pb)$  & 0.23 & 0.22 & 0.22 \\

\hline & \multicolumn{3}{c|}{$pp\rightarrow\widetilde{\chi}_{1}^{+}\widetilde{\chi}_{1}^{0}$}\\
\cline{2-4} $u\overline{d}\rightarrow\widetilde{\chi}_{1}^{+}\widetilde{\chi}_{1}^{0}$
& 0.10 & 0.09 & 0.09 \\
$\overline{d}u\rightarrow\widetilde{\chi}_{1}^{+}\widetilde{\chi}_{1}^{0}$
& 0.10 & 0.09 & 0.09 \\
\hline
$\sigma_{p_{2}}(pb)$  & 0.20 & 0.19 & 0.19 \\

\hline & \multicolumn{3}{c|}{$pp\rightarrow\widetilde{\chi}_{1}^{+}\widetilde{\chi}_{2}^{0}$}\\
\cline{2-4} $u\overline{d}\rightarrow\widetilde{\chi}_{1}^{+}\widetilde{\chi}_{2}^{0}$
& 0.13 & 0.12 & 0.12 \\
$\overline{d}u\rightarrow\widetilde{\chi}_{1}^{+}\widetilde{\chi}_{2}^{0}$
& 0.12 & 0.12 & 0.12 \\
\hline
$\sigma_{p_{3}}(pb)$  & 0.25 & 0.24 & 0.24 \\
\hline
\end{tabular}
\end{center}
\caption{The cross-sections for the chargino production  and
co-production at the LHC in the focus point region for $\mu=200$
GeV and $\tan\beta=10,30,50$\label{t200}}
\end{table}

\begin{figure}[pbht]
    \epsfxsize=7.4cm \epsffile{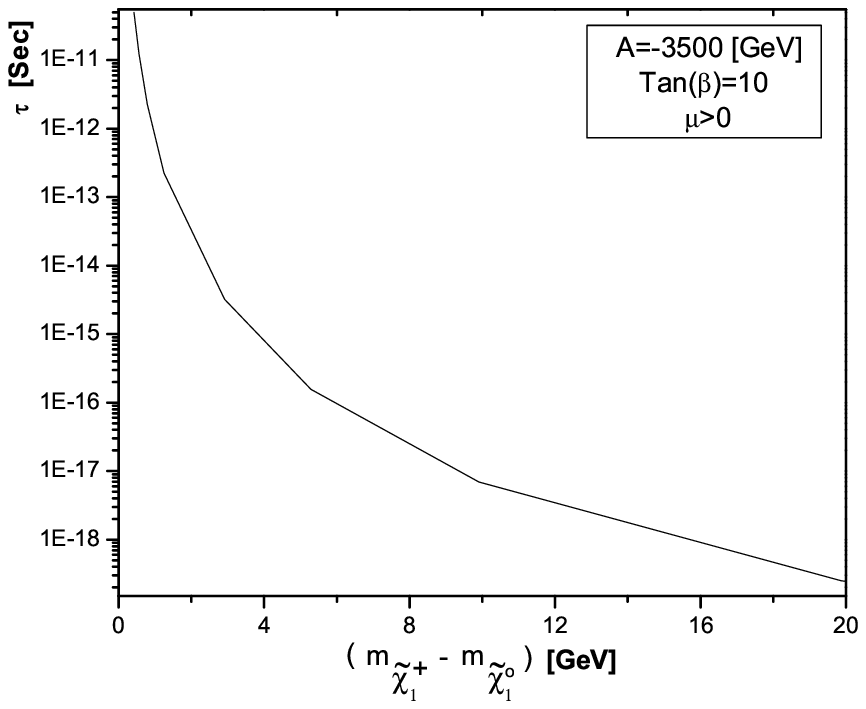}
    \epsfxsize=7.5cm\epsffile{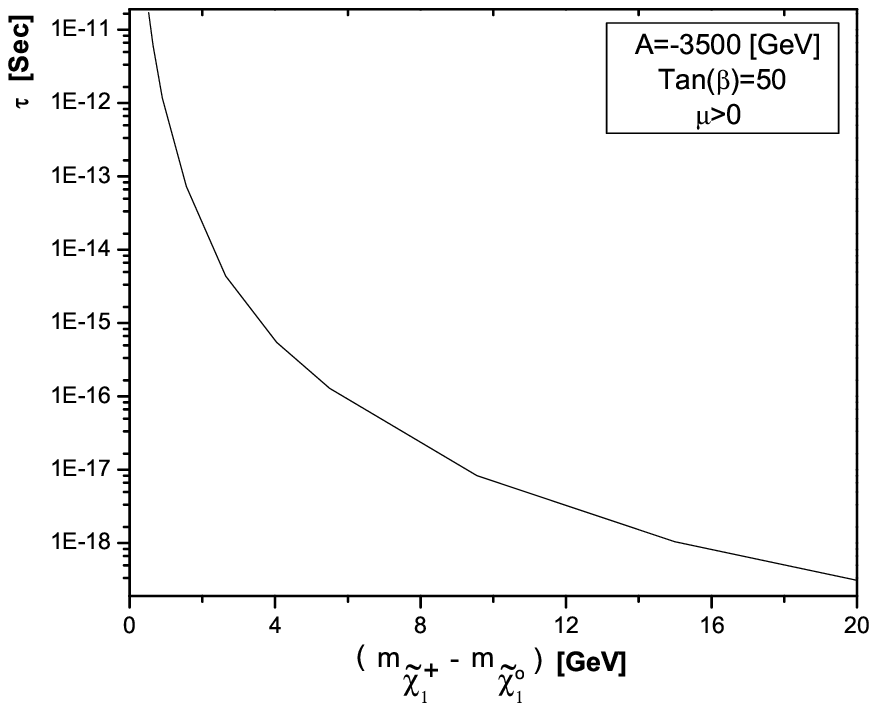}
\caption{The lightest chargino lifetimes as a function of the mass
difference between the lightest chargino (NLSP) and the lightest
neutralino (LSP).
\label{life}}
\end{figure}

In Fig.~\ref{life} we show the lifetime of the lightest  chargino
as a function of the mass difference between the lightest chargino
(NLSP) and the lightest neutralino (LSP). It appears that in order
to get reasonable "large" lifetimes one has to go very far along
the focus point region. Then keeping $\mu$ small one can get
lifetimes of the order of $10^{-10}$ s for practically degenerate
LSP and NLSP. When the mass difference increases the lifetime
falls down. However, if the degeneracy is within a few GeV,
charginos are long-lived.

\section{Conclusion}

Experimentally, of course, the possibility to have a  charged
heavy long-lived particle is appealing. As well as in the case of
long-lived stau and stop, which we considered in our previous
papers~\cite{us1,us2,us3,us4}, we may conclude that within the framework
of the MSSM with the soft supersymmetry breaking mechanism it is
possible to get long-lived superpartners of the Higgs (or
$W$-boson) which might be produced at the Large Hadronic Collider.
The production cross-sections again crucially depend on a single
parameter –- the mass of the superparticle and for light charginos
it can reach a few pb. As in the case of light stops, the light
chargino NLSP scenario requires large negative values of the soft
trilinear SUSY breaking parameter $A_0$, and large values of $m_0$
and $m_{1/2}$. The chargino decay would have an unusual signature and
produce a noticable signal rather than pure missing energy taken
away by the lightest neutralino. The two options are: charginos
go through the detector, or they produce a secondary vertex when
they decay inside the detector.

\subsection*{Acknowledgements}
Financial support from RFBR grant \# 08-02-00856-a and grant of
the Ministry of Education and Science of the Russian Federation
\# 1027.2008.2 is acknowledged.

\end{document}